\documentclass[12pt]{iopart}
\expandafter\let\csname equation*\endcsname=\relax 
\expandafter\let\csname endequation*\endcsname=\relax 
\usepackage{amsmath,amssymb,amsthm} 

\usepackage{physics}

\usepackage{subcaption}
\usepackage{xcolor}
\usepackage{float}
\usepackage{graphicx}

\begin{document}
\title[]{Novel Machine Learning and Differentiable Programming Techniques applied to the VIP-2 Underground Experiment}

\author{Fabrizio Napolitano\textsuperscript{1},
Massimiliano Bazzi\textsuperscript{1},
Mario Bragadireanu\textsuperscript{2,1},
Michael Cargnelli\textsuperscript{3},
Alberto Clozza\textsuperscript{1},
Luca De Paolis\textsuperscript{1},
Raffaele Del Grande\textsuperscript{4,1},
Carlo Fiorini\textsuperscript{5},
Carlo Guaraldo\textsuperscript{1},
Mihail Iliescu\textsuperscript{1},
Matthias Laubenstein\textsuperscript{6},
Simone Manti\textsuperscript{1},
Johann Marton\textsuperscript{3},
Marco Miliucci\textsuperscript{1,$\dag$},
Kristian Piscicchia\textsuperscript{7,1},
Alessio Porcelli\textsuperscript{7,1},
Alessandro Scordo\textsuperscript{1},
Francesco Sgaramella\textsuperscript{1},
Diana Laura Sirghi\textsuperscript{7,1,2},
Florin Sirghi\textsuperscript{1,2},
Oton Vazquez Doce\textsuperscript{1},
Johann Zmeskal\textsuperscript{3,1} and
Catalina Curceanu\textsuperscript{1,2}}


\address{
    \textsuperscript{1} INFN, Laboratori Nazionali di Frascati, Via E. Fermi 54, Frascati I-00044, RM, Italy\\
    \textsuperscript{2} IFIN-HH, Institutul National pentru Fizica si Inginerie Nucleara Horia Hulubei, Str. Atomistilor No. 407, Bucharest-Magurele, Romania\\
    \textsuperscript{3} Stefan-Meyer-Institute for Subatomic Physics, Austrian Academy of Science, Kegelgasse 27, 1030, Vienna, Austria\\
    \textsuperscript{4} Physik Department E62, Technische Universität München, James-Franck-Straße 1, 85748, Garching, Germany\\
    \textsuperscript{5} Politecnico di Milano, Dipartimento di Elettronica, Informazione e Bioingegneria and INFN Sezione di Milano, 20133, Milano, Italy\\
    \textsuperscript{6} INFN, Laboratori Nazionali del Gran Sasso, Via G. Acitelli 22, 67100, Assergi, AQ, Italy\\
    \textsuperscript{7} Centro Ricerche Enrico Fermi--Museo Storico della Fisica e Centro Studi e Ricerche ``Enrico Fermi'', Via Panisperna 89a, 00184, Roma, RM, Italy\\
    \textsuperscript{$\dag$} Current position: Italian Space Agency, Via del Politecnico, s.n.c, 00133 - Roma, RM, Italy\\
}
\ead{fabrizio.napolitano@lnf.infn.it}
\vspace{10pt}
\begin{indented}
\item[]May 2023
\end{indented}

\begin{abstract}
In this work, we present novel Machine Learning and Differentiable Programming enhanced calibration techniques used to improve the energy resolution of the Silicon Drift Detectors (SDDs) of the VIP-2 underground experiment at the Gran Sasso National Laboratory (LNGS). We achieve for the first time a Full Width at Half Maximum (FWHM) in VIP-2 below 180 eV at 8 keV, improving around 10 eV on the previous state-of-the-art. SDDs energy resolution is a key parameter in the VIP-2 experiment, which is dedicated to searches for physics beyond the standard quantum theory, targeting Pauli Exclusion Principle (PEP) violating atomic transitions. Additionally, we show that this method can correct for potential miscalibrations, requiring less fine-tuning with respect to standard methods. 

\end{abstract}

%
\vspace{2pc}
%
\submitto{\MST}
%
%
%

\section{Introduction}
\label{sec:introduction}
The Pauli Exclusion Principle (PEP) is a key ingredient of the quantum theory, and its violation, albeit tiny, could be motivated by physics beyond the Standard Model, in scenarios such as the violation of Lorentz invariance, existence of extra dimensions, and quantum gravity scenarios, as discussed in recent studies~\cite{brahma2017linking,arzano2016deformed,PhysRevD.107.026002}.
In this context, the VIP-2 experiment~\cite{shi2018experimental} at the Gran Sasso National Laboratory (LNGS) searches for signals of PEP violation in the form of anomalous X-ray transitions in copper atoms, using several arrays of state-of-the-art Silicon Drift Detectors (SDDs). 
The calibration of the SDDs is a critical experimental task, and it is performed in-situ using fluorescence $K_\alpha$ and $K_\beta$ X-rays lines from manganese and titanium, activated by a Fe-55 radioactive source, and a copper $K_\alpha$ line from the target material, activated by residual environmental radiation, around two orders of magnitude less intense.
The PEP-violating $K_\alpha$ transition in copper is expected to appear just a few hundred of eV below the standard copper $K_\alpha$ line. In view of the experimental goal, and to ensure an accurate calibrated energy spectrum with minimal uncertainty in energy scale and resolution, precise control to determine the copper $K_\alpha$ transition is required.
However, the difference in yield between the fluorescence lines and the copper line forces a trade-off between energy scale uncertainty and resolution. If the duration of the calibration batch is long enough, a good precision can be achieved on the copper line position. In this case, however, the calibration will be unable to capture small fluctuations on the SDDs response, worsening the energy resolution. Conversely, if the calibration batch is short, the fluctuations could be captured, but the uncertainty on the determination of the copper position will be high, due to the low yield, again worsening the energy resolution. 
Various peak-finder algorithms have been tested to be used for a fit to the spectral lines, but at low yield, standard methods are sensitive to statistical fluctuations, and require extensive fine-tuning. Moreover, the fit does not produce a sufficient accuracy without well-placed initialization of the parameters.
Machine learning (ML) techniques have found great success in signal processing in the last decade, producing stable results with substantially fewer calibration and tuning requirements with respect to algorithmic methods (as an example in deep underground detectors~\cite{Holl2019}). In our experiment, we trained a deep neural network on synthetic data reproducing the spectra's features, using convolutional neural networks, to predict the position of the spectroscopic lines.
Finally, the output parameters of the neural network are optimized by gradient descent within an automatic differentiation framework, where the loss function is expressed in terms of an unbinned likelihood function of the data given the spectrum's Probability Density Function (PDF). 
The application of this method demonstrates gains in the detector's resolution, additionally showing the capacity of recovering miscalibrations. 

We  introduce the experimental setup, detector and data taking in Section~\ref{sec:sdds}; in Section~\ref{sec:method} we describe the novel ML and differentiable programming approach, and in Section~\ref{sec:results} the results and discussion of the comparison with the standard approach. Section~\ref{sec:conclusions} presents the conclusions of this work.


\section{Experimental Setup}
\label{sec:sdds}

The VIP-2 experiment at the LNGS is proving world-class upper limits on the PEP violation probability in the open system scenarios~\cite{Piscicchia2020}. Protected by about 3600 m of water-equivalent shielding of the Gran Sasso d'Italia massif, and by lead and copper walls, the experimental setup employs Silicon Drift Detectors, housed in a vacuum chamber and placed around a copper target, where a strong direct current is circulated to provide electrons to test PEP. The SDDs~\cite{Lechner1996,lechner2001silicon} are ideal detectors for precision X-ray spectroscopy, having so far demonstrated a resolution of about 190 eV FWHM at 8 keV~\cite{depaolissif,napolitano2022testing}. Those used in VIP-2 were developed in a collaboration between Stefan Meyer Institute (SMI) of the Austrian Academy of Sciences, the Politecnico di Milano, INFN-LNF and Fondazione Bruno Kessler~\cite{quaglia2015silicon}. The thickness of 450 $\mu m$, an active area of 0.64 cm$^2$ and an efficiency of 99\% at 8 keV allow for great physics reach and satisfy the scientific requirements.

\begin{figure}
\centering
\includegraphics[width=13.5cm]{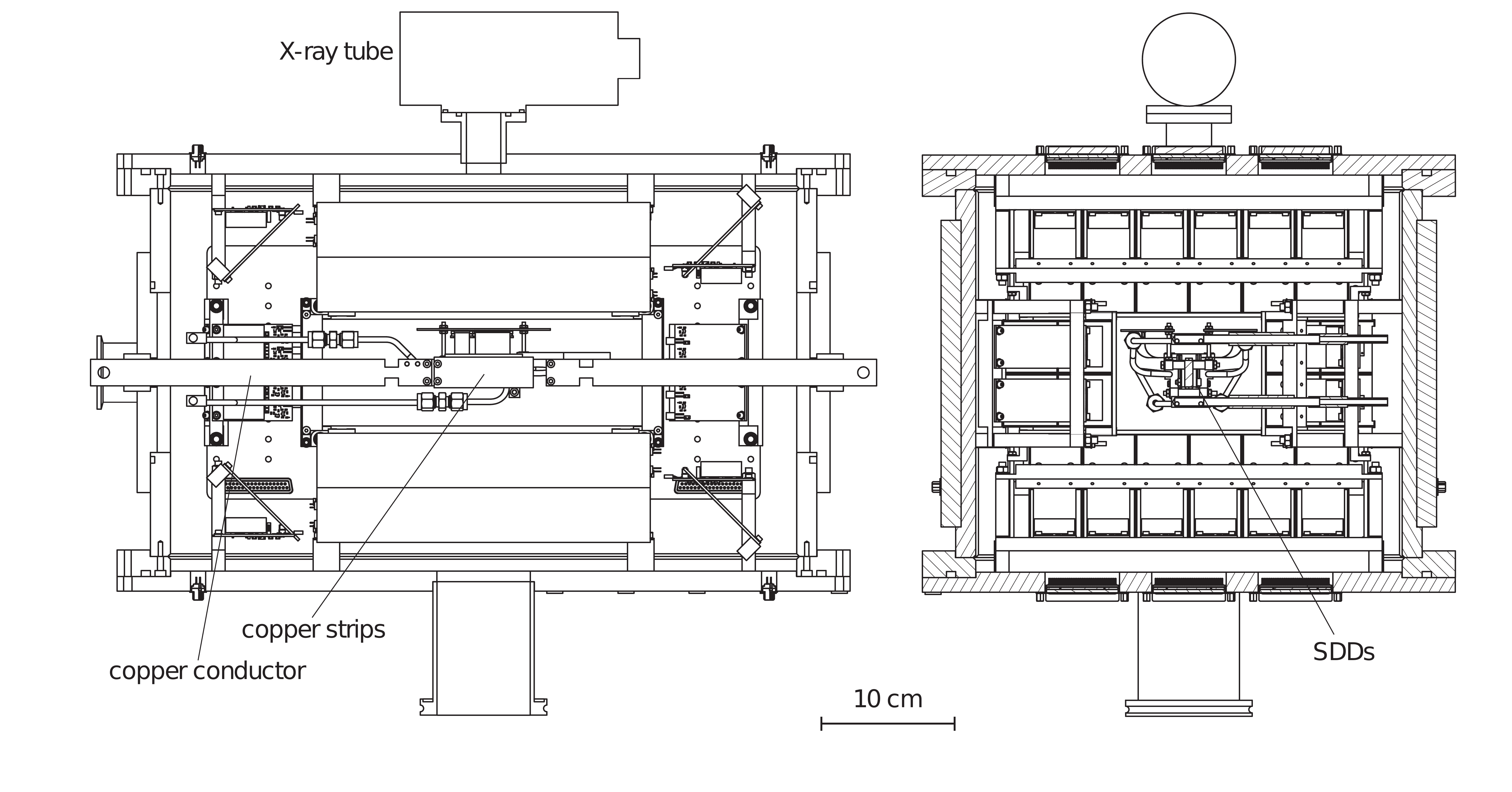}
\caption{Lateral and transverse view of the VIP-2 apparatus. The four SDDs arrays are placed in front of the target, two on one side, two on the opposite side~\cite{napolitano2022testing}.} 
\label{fig1}
\end{figure}

The four SDD arrays of VIP-2 are placed parallel to the target surface, two on one side and two on the opposite side, each array having 2 $\times$ 4 SDD cells, and being operated at a temperature of -90 Celsius. The vacuum chamber, where the SDDs and their front-end electronics are located, is operated at about $10^{-6}$ mbar.
The VIP-2 apparatus is schematically shown in Figure~\ref{fig1} with more details provided in Ref~\cite{napolitano2022testing}.

\paragraph{Data}
\begin{figure}
\centering
\includegraphics[width=11.5cm]{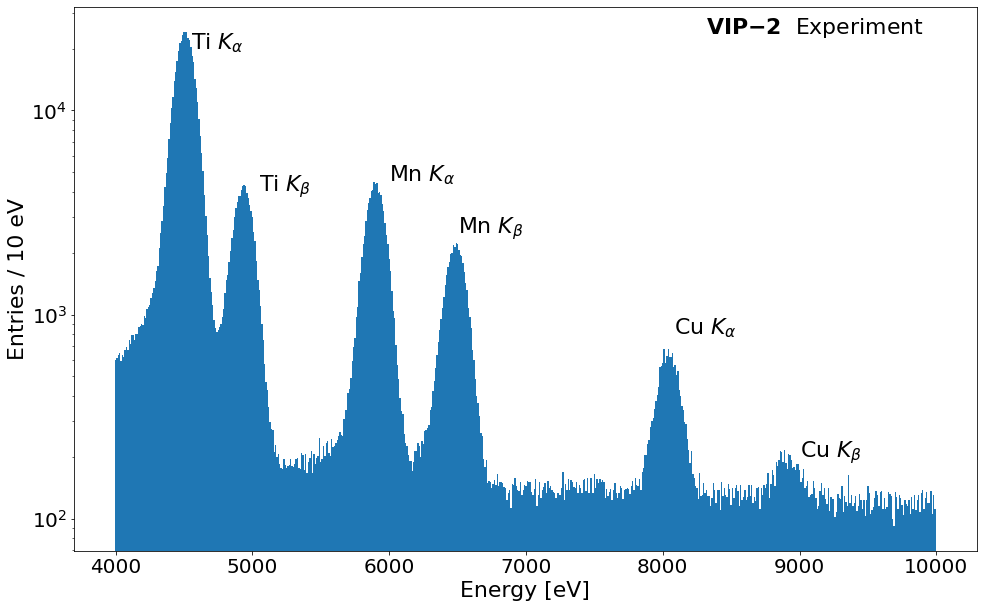}
\caption{VIP-2 energy spectrum calibrated with the standard approach in the energy range 4000-10000 eV. The spectroscopic lines are indicated on the plot. This spectrum corresponds to about 1 month of data taking, and it is used as baseline for this study. } 
\label{fig2}
\end{figure}

The data used in this work, aiming to test the novel techniques, corresponds to about a month of acquisition during 2022/2023 data taking campaign of the VIP-2 experiment. During this period, no current was circulated on the target.
The spectrum calibrated with the standard approach is shown in Figure~\ref{fig2}, where the calibration lines, Ti $K_\alpha$, Ti $K_\beta$, Mn $K_\alpha$, Mn $K_\beta$ and the copper $K_\alpha$ and $K_\beta$ lines are indicated.

The batch used for the optimization study is around 2 days of data taking, while the standard calibration batch previously used for the VIP-2 experiment is around a month.

\section{Method} 
\label{sec:method}
In Subsection~\ref{subsec:nn}, we employ a machine learning approach used to predict the centroids of the spectroscopic peaks, less sensible to statistical fluctuation than standard peak-finder algorithms (Subsection~\ref{subsec:nn}). In the following Subsection~\ref{subsec:diff} we introduce a novel approach based on differentiable programming to optimize the calibration of the SDDs, reaching below 180 eV FWHM at the copper line.

\subsection{Identification of the spectroscopic centroids at low yield}
\label{subsec:nn} 
\begin{figure}
\begin{subfigure}[b]{0.35\textwidth}
\includegraphics[clip,width=5.0cm]{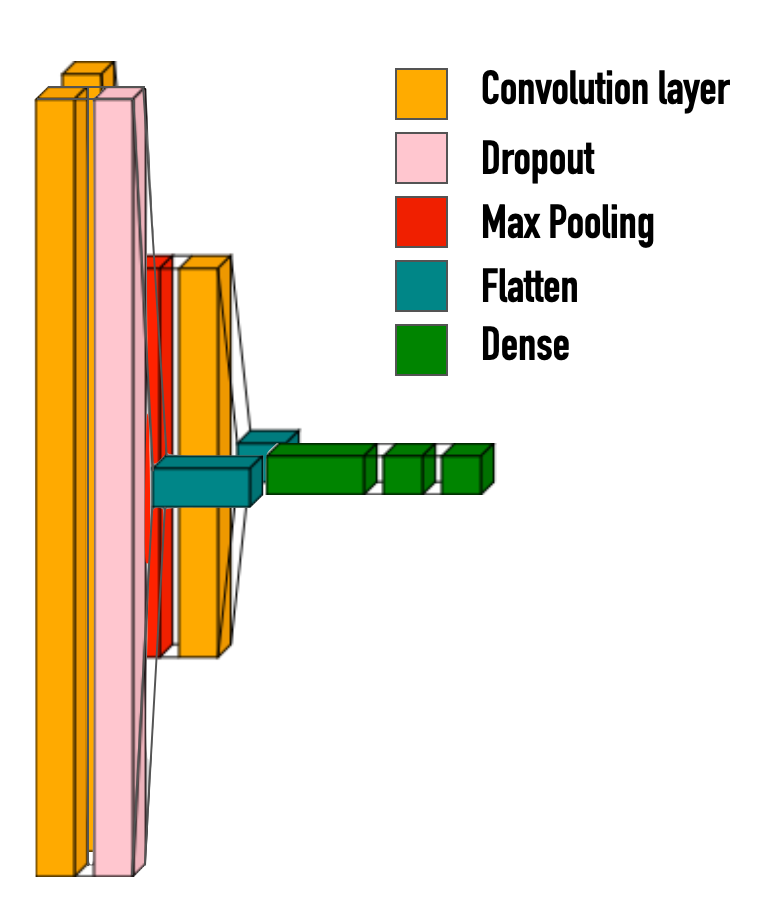}
\end{subfigure}
\begin{subfigure}[b]{0.5\textwidth}
\includegraphics[clip,width=10.0cm]{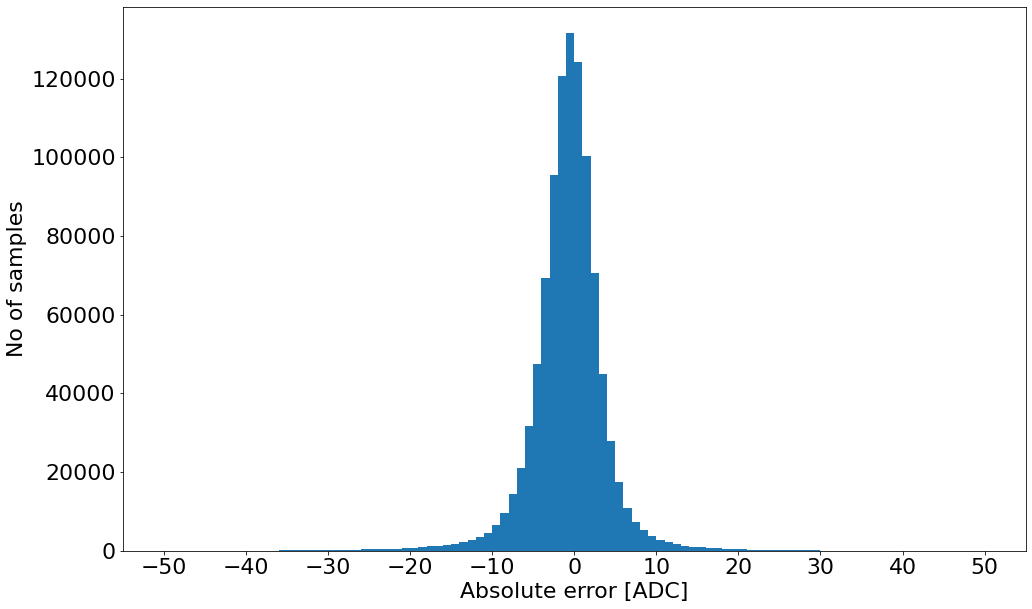}
\end{subfigure}
\caption{On the left, the schematic of the architecture employed on VIP-2 data. The input is taken from two parallel convolutional networks. The first one has kernels of bigger size, in order to learn features of higher scales, the second one has a smaller size, which instead is applied with maximum granularity. The combined output is then used as input to three layers of a fully connected network, where the output nodes are the normalized position of the calibration peaks. In the diagram, orange represents a convolutional layer, pink a dropout, red a max pooling, teal a flatten layer, and green a fully connected one. On the right, the difference between predicted and true centroid from simulated data. The result shows a good agreement within ten ADCs.}
\label{fig3}
\end{figure}

We employ a neural network architecture trained on synthetic data, reproducing the features of the real data. The model architecture is described in Figure~\ref{fig3} (left), using two convolutional towers to learn features of bigger and smaller kernel size. The model architecture is inspired from the image recognition architectures~\cite{szegedy2015going}. The synthetic data is produced with large differences in peak positions, widths and relative amplitudes, in order to stabilize the performance even in case of important fluctuations. This conservative approach assures stability of the method in a wide range of cases.

\paragraph{Network architecture}
The network is a convolutional neural network (CNN) with two branches. The input is a 1D sequence of length 300, which represent the uncalibrated energy spectrum.
The convolutional branch consists of two 1D convolutional layers with ReLU (Rectified Linear Unit) activation. The first layer has 5 filters and a kernel size of 50; the second layer has 5 filters and a kernel size of 10. Max pooling is applied with a pool size of 2.
The dense branch starts with a 1D convolutional layer with 5 filters and a kernel size of 1, followed by a dropout layer. 
The outputs of both branches are flattened and concatenated, and three dense layers with ReLU activation are applied, with 500 units, 50 units, and 5 units respectively.
This network architecture captures local and global patterns in the input data through convolutional operations, pooling, and dense layers. The final layer produces the network's output with the five node representing the position of the five spectroscopic centroids.

\begin{figure}
\begin{subfigure}[b]{0.49\textwidth}
\includegraphics[clip,width=7.9cm]{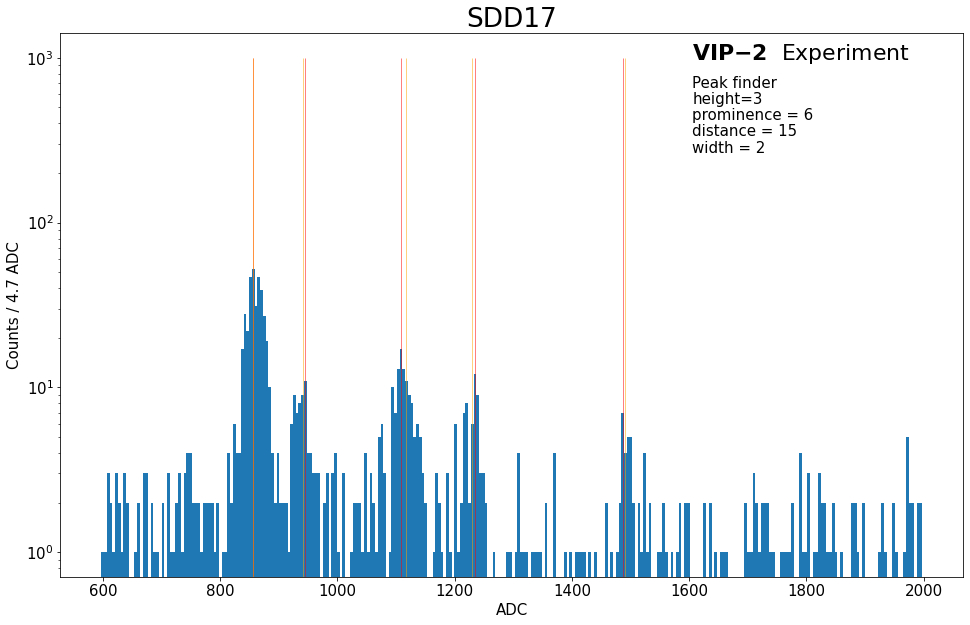}
\end{subfigure}
\begin{subfigure}[b]{0.49\textwidth}
\includegraphics[clip,width=7.9cm]{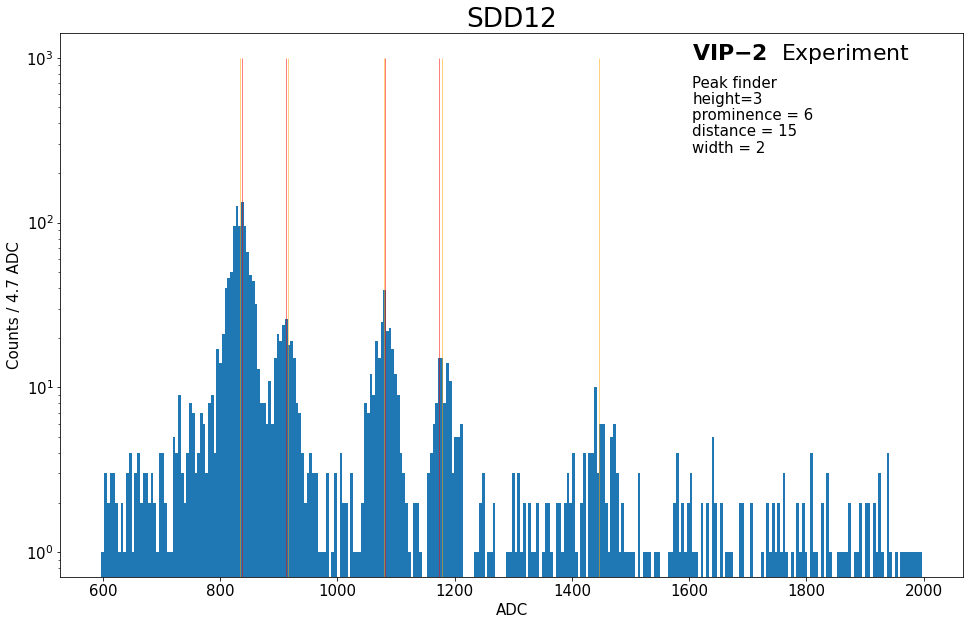}
\end{subfigure}
\\
\centering
\begin{subfigure}[b]{0.49\textwidth}
\includegraphics[clip,width=7.9cm]{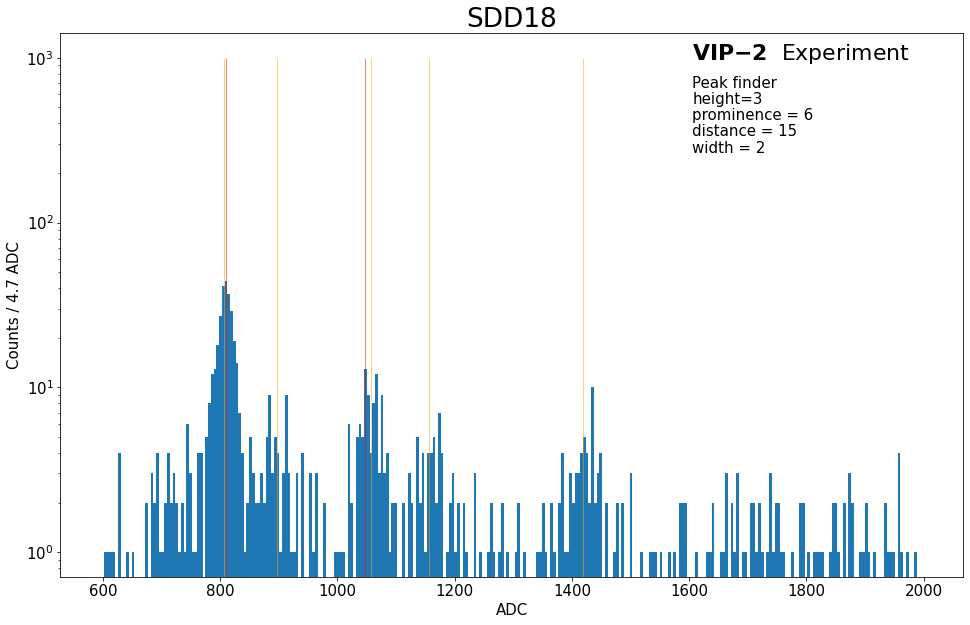}
\end{subfigure}

\caption{The ADC spectrum of selected SDDs (17, top left; 12 top right; 18 bottom) taken by the VIP-2 experiment in 2022, which exhibit a similar, yet slightly different relative yields. The orange lines are obtained by the neural network prediction, and are found to correctly identify the spectroscopic lines of interest (titanium, manganese and copper), even with low counts and statistical fluctuations in all the considered cases. The red lines represent \texttt{SciPy} peak finder algorithm, with the algorithm parameters printed on the top right of the plot. Due to the different levels of statistical fluctuations, the algorithm does not identify copper $K_\alpha$ (SDD12) and  Ti $K_\beta$, Mn $K_\beta$ (SDD18) with the same set of parameters, requiring extensive calibration and fine-tune.}
\label{fig4}
\end{figure}

\paragraph{Performance}
The trained model is subsequently fine-tuned on synthetic data, but with much lower statistics. On Figure~\ref{fig3} (right), the absolute error of the network prediction with respect to the true values used to generate the data is shown. The results are correct within ten Analogue to Digital Converter counts (ADCs). While similar performances can be obtained by state-of-the-art peak finder algorithms~\cite{du2006improved,antcheva2011root}, this method demonstrates stable performances even for low count numbers, corresponding to small integrated times, mitigating the need for extensive fine-tuning of the algorithms. In Figure~\ref{fig4}, the ADC spectrum of a single SDD acquired in 2022 by the VIP-2 experiment, together with the corresponding neural network prediction, are shown.

The predicted centroids were then used as seeds for a subsequent shape fit.
Finally, the calibration constants were derived assuming a polynomial energy response:
\begin{equation}
\text{E} =  \mathcal{C}(\text{ADC},p_0,p_1,p_2) =  p_0 + p_1 \times \text{ADC} + p_2 \times \text{ADC}^2
\end{equation}

The SDDs energy response is known to have a linear dependency on the ADC~\cite{Miliucci2019}, however a quadratic term was also introduced to correct for possible distortions at the level of the front-end electronics.
Additionally, the relatively simple form enables a differentiable programming approach, as shown in the next Subsection.

\subsection{A differentiable programming approach to calibration optimization}
\label{subsec:diff} 

Differentiable programming is a computational paradigm that enables the automatic differentiation of mathematical functions, making it a valuable tool for scientific computing and machine learning,  already pioneered in High Energy Physics~\cite{de2019inferno,simpson2023neos,https://doi.org/10.48550/arxiv.2203.13818}. Within this paradigm, it is possible to employ high-level syntax to form fully differentiable complex models that can be optimized using gradient-based methods.
The key feature of differentiable programming is its ability to automatically compute function gradients with respect to their inputs. This is achieved through a process called reverse-mode automatic differentiation, which propagates gradients backward through the function graph using the chain rule of calculus. This allows for efficient computation of gradients, even for functions with numerous inputs and outputs.

We used an approach based on differentiable programming to optimize the $p_0$, $p_1$ and $p_2$ constants for each one of the 32 SDD cells and for each calibration batch. For this purpose,  JAX~\cite{jax2018github}, an open-source numerical computing library designed for high-performance machine learning research, was used.

Since we are dealing with three calibration constants for each $i\in 1\dots N$ calibration batch, we define the calibration matrix as:
\begin{equation}
\mathbf{P} = \begin{bmatrix}
p_{0,1} & p_{1,1}& p_{2,1}\\
\vdots & \vdots & \vdots\\
p_{0,i} & p_{1,i}& p_{2,i}\\
\vdots & \vdots & \vdots\\
p_{0,N} & p_{1,N}& p_{2,N}\\
\end{bmatrix}
\end{equation}

The goal is to optimize this matrix to get a better spectroscopic response of the SDDs. In order to do that, leveraging differentiable programming, we need to define a loss-function. We define an unbinned likelihood-based loss-function, taking into account the shape of the copper $K_\alpha$ line, since it is the closest to the PEP violating Region of Interest (7000-8500 eV):

\begin{multline}\label{eq1}
\mathcal{L}(\mathbf{ADC},\mathbf{P})  =\prod_i \prod_{j\in i} \big( \text{Gauss} (\mathcal{C}(\text{ADC}_{i,j},{P}_i) - \mu_{Cu_{K_{\alpha 1}}},\sigma)  \\ +
 \text{Gauss} (\mathcal{C}(\text{ADC}_{i,j},{P}_i) - \mu_{Cu_{K_{\alpha 2}}},\sigma) \big) 
\end{multline}
where $\text{ADC}_{i,j}$ is the $j$-th value of the ADC in the batch $i$, Gauss is the Gaussian distribution, $\mu_{Cu_{K_{\alpha 1}}} = 8047.8$ eV, $\mu_{Cu_{K_{\alpha 2}}} = 8027.8$ eV ~\cite{krause1979natural,bearden1967x}, and $\sigma$ the width.
$P_i$ is the set of calibration constants $\{p_{0,i},p_{1,i},p_{2,i}\}$ in the $i$-th batch.
The gradient of this function with respect to the calibration parameters $\mathbf{P}$ is computed with JAX, allowing the optimization. For enhanced numerical and convergence stability, the logarithm of the likelihood is used.

\begin{figure}
\centering
\includegraphics[width=11.5cm]{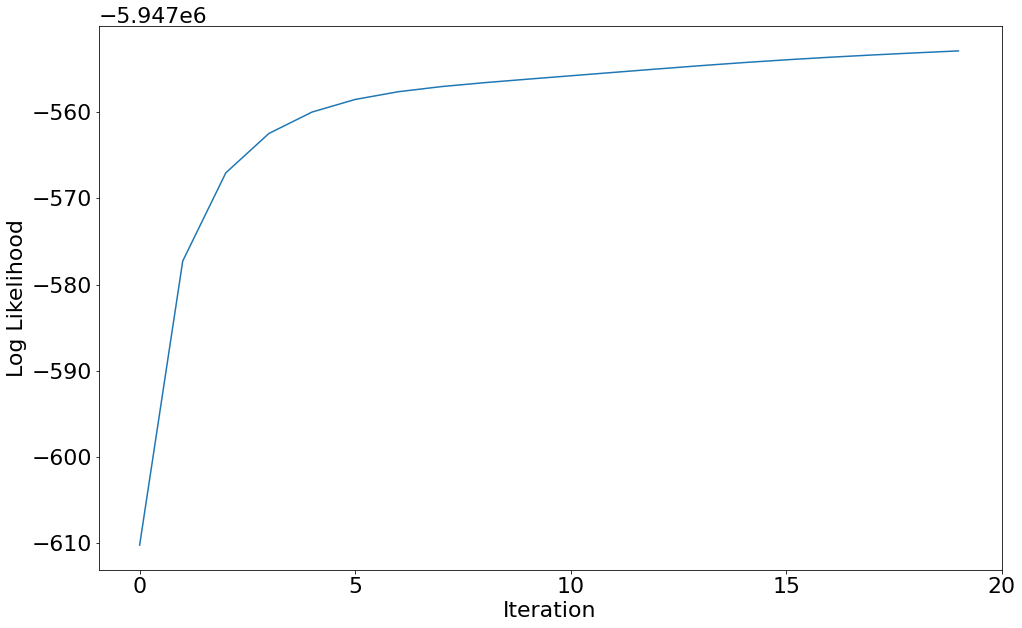}
\caption{The value of the Log Likelihood defined in Equation~\ref{eq1} as a function of the number of iterations in the gradient-descent. } 
\label{fig5}
\end{figure}
We find that already after a few iterations, the optimization converges to a maximum, as shown in Figure~\ref{fig5}.

Additionally, as a cross- and sanity-check, we quantified how much the optimized calibration parameters differ from the reference, by studying 
 $(\mu_{Cu}^{Ref}-\mu_{Cu}^{Opt})/\sigma\mu_{Cu}^{Ref}$,
 where $\mu_{Cu}^{Ref}$ is the reference Cu $K_\alpha$ centroid position with statistical error $\sigma\mu_{Cu}^{Ref}$, and $\mu_{Cu}^{Opt}$ is the one corresponding to the optimized parameters. The statistical error is obtained from the fit in MINUIT~\cite{antcheva2011root}.

\begin{figure}
\centering
\includegraphics[width=11.5cm]{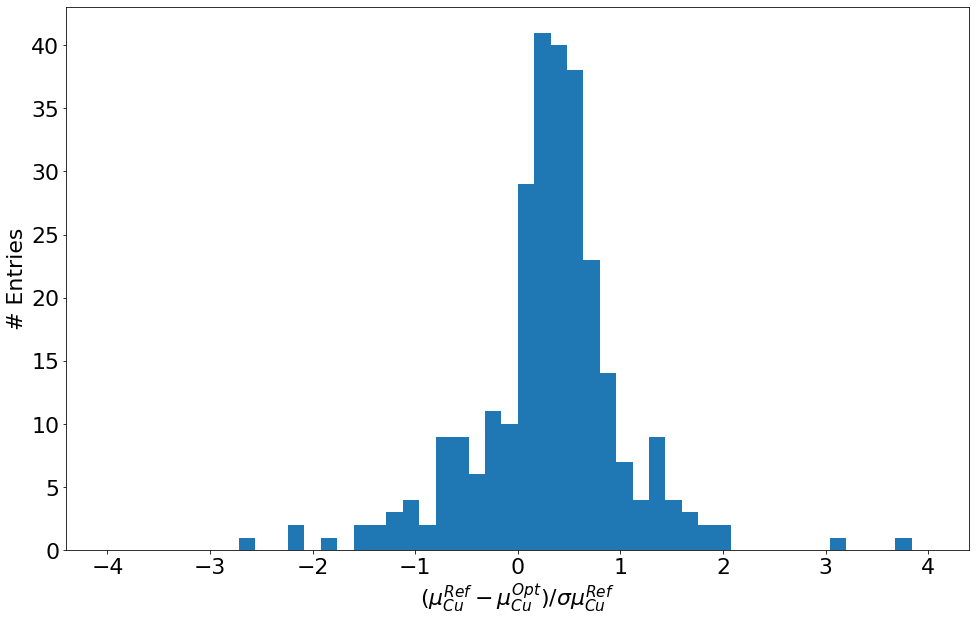}
\caption{The relative difference of the Cu $K_\alpha$ positions before and after gradient descent, normalized with respect to its statistical error. } 
\label{fig51}
\end{figure}

The histogram of these relative differences is shown in Figure~\ref{fig51}, where it can be seen that the vast majority of the centroid displacements following the optimization procedure are well within the  1$\sigma$ statistical error of the fit, ensuring that the optimization is not significantly changing the calibration parameters, as expected. 

\section{Results \& Discussion} 
\label{sec:results}

\begin{figure}[H]
\centering
\includegraphics[width=11.5cm]{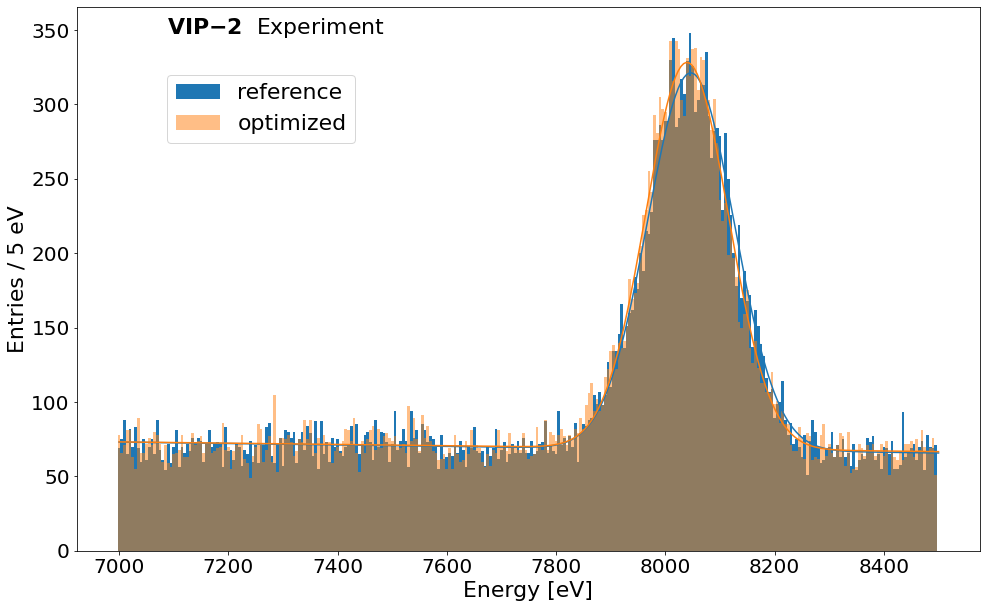}
\caption{The copper line before (blue) and after (orange) the optimization. The solid lines represent the best fits to the spectra, in the energy range 7000-8500 eV, which is the experimental ROI of the VIP-2 experiment.  } 
\label{fig6}
\end{figure}

In order to assess the overall gains of the method, we compare the two FWHMs of the copper line after and before the optimization. In Figure~\ref{fig6} the blue spectrum represents the reference data and the orange spectrum the data obtained with the enhanced calibration procedure. The energy range in the Figure is the Region of Interest (ROI) of the VIP-2 search for PEP violation in copper.
As it can be seen, the copper peak results to be more prominent, and with a smaller width. 
We quantify the enhancement of these spectroscopic qualities with a fit to the line shape. The best fit is shown as a blue and an orange line on the plot, and the best fit values are reported in Table~\ref{tab:comparison}, where the values of the peak position, FWHM and reduced $\chi^2$ are reported. 

\begin{table}[H]
\centering
\caption{Comparison of reference and optimized positions, FWHM, and $\chi^2/ndf$.}
\label{tab:comparison}
\begin{tabular}{|c|c|c|c|}
\hline
 & Position [eV]& FWHM [eV] & $\chi^2/ndf$ \\
\hline
Reference & 8050 $\pm$ 1 & 185 $\pm$ 2 & 1.64 \\
Optimized & 8048 $\pm$ 1 & 176 $\pm$ 2 & 1.25 \\
\hline
\end{tabular}
\end{table}

The fit function used to describe the shape is:
\begin{equation}
f(x,A,\mu,\sigma) = A\times\frac{51}{100}\times\text{Gauss}(x-\mu-20,\sigma)+ T_2(x) + A \times \text{Gauss}(x-\mu,\sigma) + T_1(x) + m \times x+ C  
\end{equation}
where the first term describes the $K_{\alpha 2}$ with relative intensity 51/100 and the second term describes the $K_{\alpha 1}$. 20 eV is the energy difference between the two, and the continuum background is found to be best described with a linear function.
The two contributions $T_1(x)$ and $T_2(x)$ are the tail functions, which reproduce incomplete charge collection. They are written as:
\begin{equation}
    T_i(x) = \frac{A_i}{2\beta\sigma}\times e^{ \frac{x - \nu}{\beta \sigma}\frac{1}{2\beta^2}} \times erfc \left( \frac{x-\nu}{\sqrt{2}\pi } + \frac{1}{\sqrt{2}\beta}\right)
\end{equation}
where $A_i$ is the amplitude of the tail, $\beta$ is the tail's slope, $erfc$ is the complementary error function. $\nu$ is $\mu$ for $i=1$ and is $\mu-20$ for $i=2$.
The results of the fit show that the ML and differentiable programming approach has yielded gains in each of the examined parameters. In particular, the peak position in the optimized procedure has shown to correct for the residual miscalibration, and the centroid value is now more compatible with tabulated 8047.8 eV. The FWHM of the line has reached for the first time in VIP-2 below 180 eV, showing a substantial improvement with respect to traditional methods used in the past. Finally, the reduced chi-squared is used to quantify the agreement of the data against the model. This parameter also shows an improvement, and the data are found to be more compatible with the model.

\subsection{Discussion}
Following this study applied to the most recent VIP-2 data, we outline the advantages this method brings with respect to the physics reach of the VIP-2 experiment.
First, in the analysis of the 2020 VIP-2 data~\cite{napolitano2022testing}, the energy scale uncertainty was included to reflect the differences in scale between calibrated spectrum and standard lines, originating from the calibration procedure. With the new approach, and its capability to correct for potential miscalibration, we show that this effect will be substantially reduced. 
Second, the reduced chi-squared shows that the data calibrated with this method have a higher degree of compatibility with the model, ensuring a more coherent description of the background and of the spectroscopic lines. The PEP signal description at 7.7 keV relies on the accurate knowledge of the continuum background, and the energy scale.
Finally, the better energy resolution in the form of a smaller FWHM assures a higher discovery significance of the signal, since the PEP-violating line would appear more prominently above the background.

\section{Conclusions}
\label{sec:conclusions}

The VIP-2 experiment at LNGS is searching for signals beyond the standard quantum theory, namely  Pauli Exclusion Principle forbidden transitions in copper, possibly connecting Lorentz invariance, extra dimensions and quantum gravity to atomic X-ray phenomena.
The Silicon Drift Detectors are optimal tools to inspect this energy range. However, so far their energy resolution in VIP-2 has typically never gone substantially below 190 eV. We have applied in this work an approach to leverage the calibration procedure in order to push the FWHM of the copper line at the hardware limit.
We have trained a neural network with synthetic data as peak finder to exploit the peculiar spectrum and in-situ calibration lines even at low yields, enabling the use of much smaller calibration batches.
The network output was then used as input in a fully differentiable framework, where the loss function was described in terms of the agreement with the copper line as an unbinned likelihood.
The gradient descent has shown substantial improvements of the spectroscopic properties of the detectors, both in terms of the compatibility of the line with its two components, and by bringing for the first time in VIP-2 the SDD's FWHM at 8 keV below 180 eV.
Last, but not least, the method has shown to correct for small miscalibrations.
This novel approach has the potential to improve the physics capabilities with enhanced calibration.
For VIP-2, this translates to several advantages: a smaller energy scale uncertainty, better control over the background in the ROI, and consequently a higher discovery significance for PEP violating signals.
We plan to employ this method on the entire VIP-2 dataset. Furthermore, we plan to explore the advantages it yields in experiments which strongly rely on the precise determination of spectroscopic lines in the X-ray domain, such as the SIDDHARTA-2 experiment~\cite{Miliucci2021} at the DA$\Phi$NE collider.

\section*{Acknowledgements}
We thank: the INFN Institute, for supporting the research presented in this article and, in particular, the Gran Sasso underground laboratory of INFN, INFN-LNGS, and its Director, Ezio Previtali, and the LNGS staff. We thank C. Capoccia from LNF  and H. Schneider, L. Stohwasser, and D. Stückler from Stefan-Meyer-Institut for their fundamental contribution in designing and building the VIP-2 setup.
\section*{Funding}
This research was funded by the National Institute for Nuclear Physics (INFN, Italy). This research was funded in whole, or in part, by the Austrian Science Fund (FWF) grants P25529-N20, project P 30635-N36, and W1252-N27 (doctoral college particles and interactions). This publication was also made possible through the support of Grant 62099 from the John Templeton Foundation. The opinions expressed in this publication are those of the authors and do not necessarily reflect the views of the John Templeton Foundation. We also thank the support of the H2020 FET project TEQ with grant 766900. We gratefully acknowledge support from Centro Ricerche Enrico Fermi (“Problemi aperti nella meccanica quantistica” project), and from Foundational Questions Institute (Grants No. FQXi-RFP-CPW-2008 and FQXi-MGA-2102).










\section*{Bibliography}
\bibliographystyle{unsrt}
\bibliography{main.bib}

\end{document}